\documentclass[letterpaper]{mn2e}
\usepackage{graphicx,times}

\def \kms {${\rm{km}\,\rm{s}^{-1}}$}

\def \apjs{ApJS}

\def \aanda{A\&A}
\def \apj{ApJ}
\def \aj{AJ}
\def \mnras{MNRAS}

\title[Environmental quenching: Evidence from Coma]{Environmental quenching and hierarchical cluster assembly:  Evidence from spectroscopic ages of red-sequence galaxies in Coma}
\author[Russell J. Smith et al.]
{Russell J. Smith$^{1}$\thanks{Email: russell.smith@durham.ac.uk}, 
John R. Lucey$^{1}$, 
James Price$^{2}$,
Michael J. Hudson$^{3,4}$
\newauthor and
Steven Phillipps$^{2}$\\
~\\
$^1$Department of Physics, University of Durham, Durham DH1 3LE\\
$^2$Astrophysics Group, H. H. Wills Physics Laboratory, University of Bristol, Tyndall Avenue, Bristol BS8 1TL\\
$^3$Department of Physics and Astronomy, University of Waterloo, 200 University Avenue West, Waterloo, Ontario N2L 3G1, Canada\\
$^4$Perimeter Institute for Theoretical Physics, 31 Caroline Street North, Waterloo, Ontario, N2L 2Y5, Canada\\
}

\date{Accepted 2011 October 5. Received 2011 October 4; in original form 2011 August 15}

\pagerange{\pageref{firstpage}$-$\pageref{lastpage}} \pubyear{2011}

\begin{document}

\label{firstpage}

\maketitle

\begin{abstract}
We explore the variation in stellar population ages for Coma cluster galaxies as a function of projected cluster-centric distance,
using a sample of 362 red-sequence galaxies with high signal-to-noise spectroscopy. The sample spans a wide range in luminosity (0.02--4\,$L^*$)
and extends from the cluster core to near the virial radius. 
We find a clear distinction in the observed trends of the giant and dwarf galaxies. 
The ages of red-sequence giants are primarily determined by galaxy mass, 
whether parametrized by velocity dispersion, luminosity, stellar mass or dynamical mass, with only weak modulation by
environment, in the sense that galaxies at larger cluster-centric distance are slightly younger. 
For red-sequence dwarfs (with mass $\la10^{10}$\,M$_\odot$), the roles of mass and environment as predictors of age are reversed: 
there is little dependence on mass, but strong trends with projected cluster-centric radius are observed. The average age of dwarfs at the 2.5\,Mpc
limit of our sample is approximately half that of dwarfs near the cluster centre. 
The gradient in dwarf galaxy ages is a global cluster-centric trend, and is not driven by the ongoing merger of the NGC 4839 group to the south west of Coma.
We interpret these results using environmental histories extracted from the {\it Millennium Simulation} for members of massive clusters.
Hierarchical cluster assembly naturally leads to trends in the accretion times of galaxies as a function of projected cluster-centric radius. 
On average, simulated galaxies now located in cluster cores joined halos above any given mass threshold earlier than those now in the
outskirts of clusters. We test environmental quenching models, in which star formation is halted in galaxies when they enter halos of
a given mass, or when they become satellites. Our models broadly reproduce the gradients observed in Coma,
but for dwarf galaxies the efficiency of environmental quenching must be very high to match the strong trends observed.
\end{abstract}

\renewcommand{\textfraction}{0.5}

\section{Introduction}\label{sec:intro}

According to the current cosmological paradigm, galaxy formation and evolution takes place within a hierarchy of structure formation 
governed by the merging of dark-matter halos. The richest galaxy clusters at $z=0$, with total masses of $\ga10^{15}$\,M$_\odot$
occupy the apex of this hierarchy. By definition, such clusters are highly unrepresentative of the present-epoch Universe. 
Moreover, the galaxies now collected within such clusters formed and evolved in regions of above-average density at {\it all} epochs of their history, 
leading to evolution that was accelerated compared to those in more representative regions. 

The evolution of galaxies is linked to their surroundings through a rich variety of environmental processes (e.g.  Boselli \& Gavazzi 2006, and references therein).
Most dramatically, galaxies may merge with others of similar mass, profoundly affecting their structure and triggering massive star-bursts which consume their
gas supply on short time-scales. Mergers with less massive galaxies are individually less transformative, but cumulatively may dominate the assembly of stellar mass 
in some galaxies. Even where mergers do not occur, fly-by tidal interactions with other galaxies, or the larger scale gravitational potential, can re-distribute stars 
leading to morphological transformation, or perhaps cause tidal instabilities leading to gas being shocked into star formation.
%
Galaxies can also be influenced by interaction with their gaseous surroundings:
rapid motion through a dense intra-cluster
medium can lead to ram-pressure stripping of cold interstellar gas in disks as first discussed by Gunn \& Gott (1972), or else the reservoir of hot halo
gas may be stripped, eventually starving the disk of continued gas supply (Larson, Tinsley \& Caldwell 1980; Balogh, Navarro \& Morris 2000). 
Qualitatively, the end result of these processes is to exhaust or remove the gas from galaxies in denser environments, leading to a shutdown in star formation 
(``quenching''), as required to explain population correlations such as the morphology--density relation (e.g. Dressler 1980). At a more detailed level, 
however, the relative importance of the various processes, and how the contributions depend on local and global environment, and vary over cosmic
time, remain crucial areas of uncertainty. 

Despite being unrepresentative regions by definition, rich clusters are important because they
provide observationally convenient galaxy samples for intensive study of galaxy properties, and also because {\it some} of the possible quenching mechanisms,
such as ram-pressure stripping of cold gas, are probably effective {\it only} in the densest, most massive systems. Galaxies with stripped tails directed away from 
the centres of clusters provide evidence for ongoing stripping in massive clusters (e.g. Chung et al. 2007; Sun et al. 2010; Smith et al. 2010), showing
that clusters are not merely passive recipients of dead galaxies. 
On the other hand, since many galaxies spend much of their histories in group-like environments, it is likely that these, rather than clusters, are dominant
in driving evolution in the galaxy population globally, and that many cluster galaxies were ``pre-processed'' in groups prior to accretion into their current environment
(Zabludoff \& Mulchaey 1998). 
Hence in clusters, we are likely observing the remnants of quenching processes that  occurred in a range of environments, over a range of epochs, and
subsequently mixed as the system assembles. 

In this paper, we investigate the characteristic stellar ages of red-sequence (i.e. quenched) galaxies as a function of their location within a single 
very massive low-redshift cluster, Coma. Unlike ``instantaneous'' measures of galaxy properties, such as current star-formation rates,  or 
membership of the red sequence or blue cloud, the stellar ages in principle probe the {\it history} of quenching in the galaxy population. 
Following previous spectroscopic work which suggested a strong spatial dependence of stellar populations in Coma  galaxies
(e.g. Guzm\'an et al. 1992; Caldwell et al. 1993; Carter et al. 2002), 
we conducted a high-signal-to-noise study of faint red-sequence galaxies in this cluster. 
Earlier results from our work (Smith et al. 2008, 2009a, hereafter S09) were based on observations in the core and in a region to the south west, where the NGC 4839 group is merging 
with the main cluster. We confirmed an age--radius relation for dwarfs which was much steeper than that recovered for more massive galaxies, in an ensemble of clusters, by Smith et al. (2006). 
However, the S09 analysis suffered from its limited spatial coverage, and in particular could not distinguish between a global age-versus-radius correlation and a 
localised excess of young galaxies associated with the sub-cluster merger. 
Studies of other clusters have provided additional hints for strong environmental trends for dwarfs (Michielsen et al. 2008; Chilingarian et al. 2008), 
which lend support to the former interpretation, but are based on samples of limited size and radial extent. 
Moreover, because S09 concentrated only on the dwarfs, the mass-dependence of the environmental trends in Coma was not uniformly investigated. 

In this paper, we extend the work of S09 using deep MMT observations for an enlarged sample of dwarfs, including additional outer
fields which remove the bias toward the south-west region. Furthermore, where S09 used data for galaxies in a different location
(the Shapley Supercluster) to provide the high-luminosity comparison sample, here we use spectra from the Sloan Digital Sky Survey
which, from Data Release 7 (Abazajian et al. 2009), covers the whole Coma region. 
The treatment of the SDSS data has been described by Price et al. (2011). Together, the SDSS and 
MMT datasets cover a six-magnitude range in luminosity, and cover most of the area inside the nominal virial radius of the cluster, enabling 
a comprehensive re-analysis of the cluster-centric dependence of passive galaxy ages. 

We interpret the observational results explicitly in the context of hierarchical structure formation, through comparison with 
cluster assembly histories drawn from the  {\it Millennium Simulation} (Springel et al. 2005). Specifically, we test models in which the 
cessation of star formation is linked to incorporation of galaxies into halos above a given threshold, or change of status from central galaxy to satellite within a halo. 
Such models lead generically to halo-centric gradients in stellar population age, since dynamical friction
causes the earliest-accreted sub-halos to sink towards the cluster centre, while the most recently added structures reside at larger radius (e.g. Gao et al. 2004).
We determine how the strength of the predicted gradients depends on the details of the adopted quenching criteria and compare the predictions with our
measurements in Coma. 

The remainder of the paper is organised as follows: 
Section~\ref{sec:obsres} covers the observational results, including a summary of the sample and data employed (Section~\ref{sec:data})
and an investigation of the cluster-centric dependence of galaxy ages (Section~\ref{sec:envtrends}).
Section~\ref{sec:modres} presents the modelling results, starting by exploring how the typical accretion history of cluster galaxies 
relates to their eventual location within the cluster halo (Section~\ref{sec:modenv}), then using this as a basis for environmental quenching
models (Section~\ref{sec:quenching}). The predictions are confronted with the measured trends in Section~\ref{sec:confront}, and some caveats
and limitations to the analysis are noted in Section~\ref{sec:caveats}.
We conclude with a  discussion of the results in the context of previous work, in Section~\ref{sec:discuss}. 

In converting to physical units, we adopt a distance of 100\,Mpc for Coma (i.e. $h=0.72$), so that one degree corresponds to 1.74\,Mpc. For reference, the virial
radius of Coma has been estimated to be 2.9\,Mpc, and its mass within this radius is $\sim1.4\times10^{15}$\,M$_\odot$ (\L{}okas \& Mamon 2003).

\section{Coma observations and age correlations}\label{sec:obsres}

\subsection{Sample, data and parameter measurements}\label{sec:data}

\begin{figure*}
\includegraphics[angle=0,width=180mm]{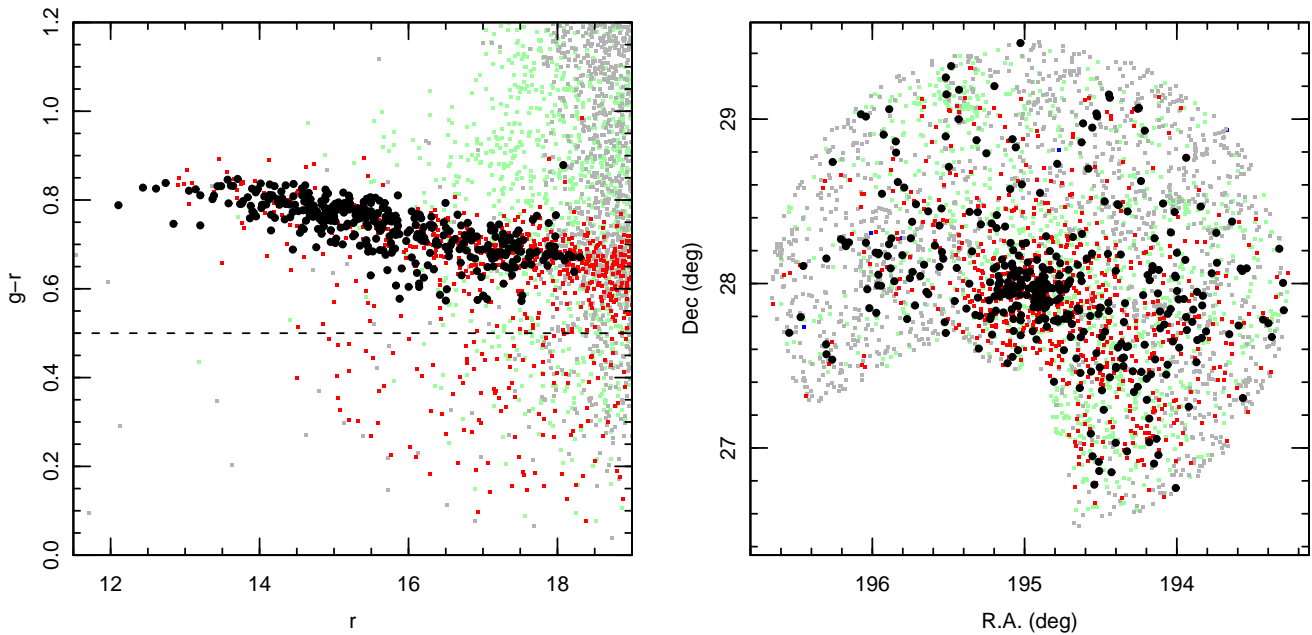}
\caption{The combined SDSS and MMT stellar populations dataset as analysed in this paper (black points), in colour--magnitude space (left) 
and projected on the sky (right). Photometry and positions are from SDSS DR7 as used in the sample construction. 
(The galaxy at $r\approx18, g-r\approx0.9$ is affected by a poorly-deblended star in DR7; its corrected colour in DR8 is 0.70).  
For comparison, we show all other SDSS galaxies within
the same spatial region, as small points coloured according to their cluster membership status: red points are spectroscopically-confirmed cluster members (with 
$3000<cz<11000$\,km\,s$^{\ \ -1}$), green points are confirmed non-members and grey points are galaxies with unknown redshift. 
Membership information includes data from our MMT/Hectospec redshift survey (Marzke et al. in preparation). The dashed line in the 
colour--magnitude diagram shows the colour cut of $g-r>0.5$ used in defining the stellar populations sample. }
\label{fig:har_sample}
\end{figure*}

Broadly the overall sample is based on spectra for 463 bright ($\sim$0.2--4.0$L^*$) Coma cluster members from SDSS (analysed in Price et al. 2011)
and 198 fainter members ($\sim$0.02--0.20$L^*$) observed using long integrations  with the Hectospec multi-fibre spectrograph at the 
6.5m MMT (Fabricant et al. 2005).
A detailed description of the sample construction, parameter measurements and methods, and data presentation, will be presented in a separate paper 
(Smith et al. 2011, hereafter S11). 
S11 will also provide a full analysis of the scaling relations of metallicity and element abundance ratios. 

Our MMT dataset is an extended version of that reported in S09. In that paper, the data were drawn from two 1-degree diameter Hectospec fields, 
one centred on the cluster core and the other displaced nearly a degree south-west to 
sample the region around the NGC 4839 subcluster. As discussed above, this limited coverage impeded a clear interpretation of the results. 
To address this, we subsequently observed a further four outer fields in a hexagonal pattern around the core (west, north-west, north-east and east), to 
yield almost complete coverage of the region within the cluster virial radius (we were unable to observe the final planned field to the south-east). 

Our Hectospec observations employed the 270\,line\,mm$^{-1}$ grating, which delivers spectra with wide wavelength coverage (3700--9000\,\AA) at a 
resolution of 4.5\,\AA\ FWHM. In each run, the high-S/N spectra required for stellar populations analysis were obtained in parallel with a 
redshift survey aimed at establishing cluster members to much deeper magnitude limits (Marzke et al. in preparation). Many fibre configurations 
were observed, each with one hour of integration, but fibres were repeatedly allocated to galaxies in the  stellar populations sample, in order to build up total 
exposure times of 4--8\,hr per object. 
The galaxy sample for the stellar populations study was drawn from known cluster members with luminosity  2--4\,mag fainter than $M^*$. 
We took into account optical colours to select passive galaxies, and SDSS spectra where available, 
to exclude galaxies with emission at H$\alpha$ (since this indicates nebular contamination of the higher-order Balmer lines used for age determination). 
The spectra were reduced as described in S09. Data from all runs were reprocessed together, to ensure that the S09  data
and the later observations were treated identically in the reduction stages. 
Error spectra were computed alongside the combined spectrum for each galaxy, for use in estimating errors on the measured parameters. 

As a complement to the Hectospec observations at low luminosity, we performed a broadly similar analysis of the 
available spectra from SDSS DR7, covering approximately the same region in the Coma cluster. The treatment of the SDSS 
data is described in detail by Price et al. (2011). For building the sample used in this paper, we relaxed the magnitude
and S/N limits imposed in the Price et al. analysis, so that we analyse all available SDSS spectra 
within two degrees radius of the Coma cluster. (We later impose a selection on age-error, 
to remove low-S/N measurements consistently across both data sources.)

The combined MMT and SDSS sample has been homogeneously re-analysed, to measure
absorption line-strength indices and emission line equivalent widths.
Stellar population ages ($T_{\rm spec}$), along with metallicities (Fe/H) and abundance ratios (Mg/Fe, Ca/Fe, C/Fe and N/Fe),
were measured via comparison of the index data against the Schiavon  (2007) simple stellar population (SSP) models, using a new model inversion code described in 
S11. The measured quantity is strictly an SSP-equivalent age, i.e. the age of a single burst which best reproduces the observed indices.
When comparing to model predictions in Section~\ref{sec:confront}, we explicitly take into account the relationship between $T_{\rm spec}$ and the true (non-SSP)
star-formation history. 
To complement the line-strength data, velocity dispersions were compiled from SDSS DR7 and a variety of literature sources
(mainly J\o rgensen 1999; Moore et al. 2002; Smith et al. 2004; Matkovi\'c \& Guzm\'an 2005;  Cody et al. 2009),
supplemented with new observations from VLT/FLAMES, and combined using observations from multiple data-sources to determine 
relative systematic offsets. Seeing-corrected half-light radii were measured from the 
$r$-band SDSS images, taking into account improved background estimation methods. The radii are used for
applying aperture corrections, and for deriving approximate dynamical masses.

There are a total of 198 galaxies in the MMT sample and 463 in the SDSS sample, but some are common to both sets, 
some (especially in the SDSS set) have strong emission lines which contaminate the stellar Balmer absorption, and 
some have insufficient signal-to-noise for reliable stellar population analysis. 
We define the sample for analysis in this paper by requiring less than a factor-of-two error in the derived age, and an equivalent width
of less than one Angstrom in emission at H$\alpha$. Additionally, we remove four galaxies from the SDSS data set that have $g-r<0.5$, substantially
bluer than all other objects in the sample, and also one galaxy from MMT that is
much fainter than the nominal limit of $r\approx18$.
For galaxies common  to both data sources, the MMT measurement is used, since its $S/N$ is always much greater than the SDSS spectrum 
(a factor of three, on average).
These cuts reduce the sample to 411 galaxies in total, of which 169 have ages measured from MMT and 242 from SDSS. Velocity dispersion data are available
for a more restricted sample, comprising 355 galaxies, of which 120 have ages from MMT and 235 from SDSS. 

Finally, for this paper, we further restrict attention to the spatial regions of the Coma cluster that were targetted in the MMT observations, 
such that the sampling is fairly similar, as a function of magnitude, at all locations. Cutting SDSS galaxies that lie in regions without MMT coverage 
(i.e. beyond a radius 1.5$^\circ$, or beyond 0.5$^\circ$ in the missing south-east wedge 110$^\circ$--190$^\circ$, measured east from north), the sample
is reduced to 362 galaxies of which 310 have velocity dispersion measurements. 
For the final sample, the median $S/N$ is 41\,\AA$^{-1}$, and the median error in age is 0.14\,dex. Because the faint part of the sample is drawn 
from the higher-$S/N$ MMT data, the data quality does not degrade rapidly for dwarf galaxies: 
for galaxies below the median luminosity of the sample ($r_{\rm petro}>15.6$), the median $S/N$ is 42\,\AA$^{-1}$, and the median 
error in age is 0.12\,dex.
Figure~\ref{fig:har_sample} shows the colour--magnitude relation and sky distribution for the final sample.

\begin{figure*}
\includegraphics[angle=0,width=180mm]{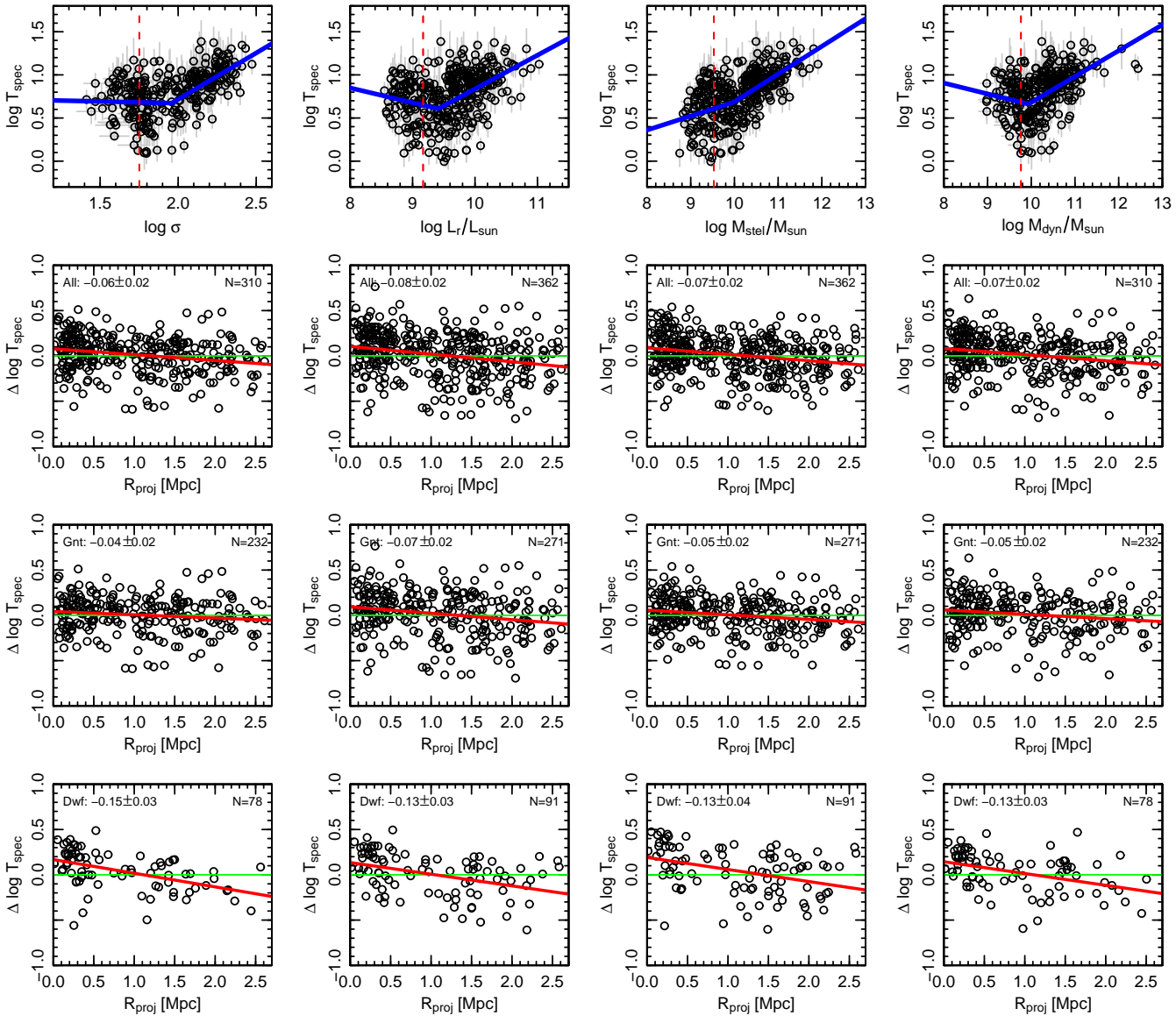}
\caption{
{\it Top row}: Age versus mass relations, using a variety of ``mass-proxy'' quantities, left to right: velocity dispersion, luminosity, stellar mass and dynamical mass. 
In each case, the heavy blue line segments show a broken-stick model-fit to the data. The vertical dashed red line 
shows the position of the lower quartile of the mass distribution. 
{\it Second row}: Residuals in age, relative to the corresponding age--mass relation in the top row, plotted against projected distance from the cluster centre, for all galaxies. The
red line shows a linear fit to the trend, with slope noted above. A zero-slope reference line is shown in green.
{\it Third row}: 
As for the second row, but showing only the galaxies in the upper three quartiles of the  mass distribution (``giants''), and a fit to this sub-sample.
{\it Bottom row}: 
Equivalent for the galaxies in the lowest quartile of the mass distribution (``dwarfs''). 
A significant trend for younger ages in the outer galaxies, especially dwarfs, is recovered with similar amplitude regardless of which mass proxy is adopted.}
\label{fig:har_agemass}
\end{figure*}

\begin{figure*}
\includegraphics[angle=0,width=180mm]{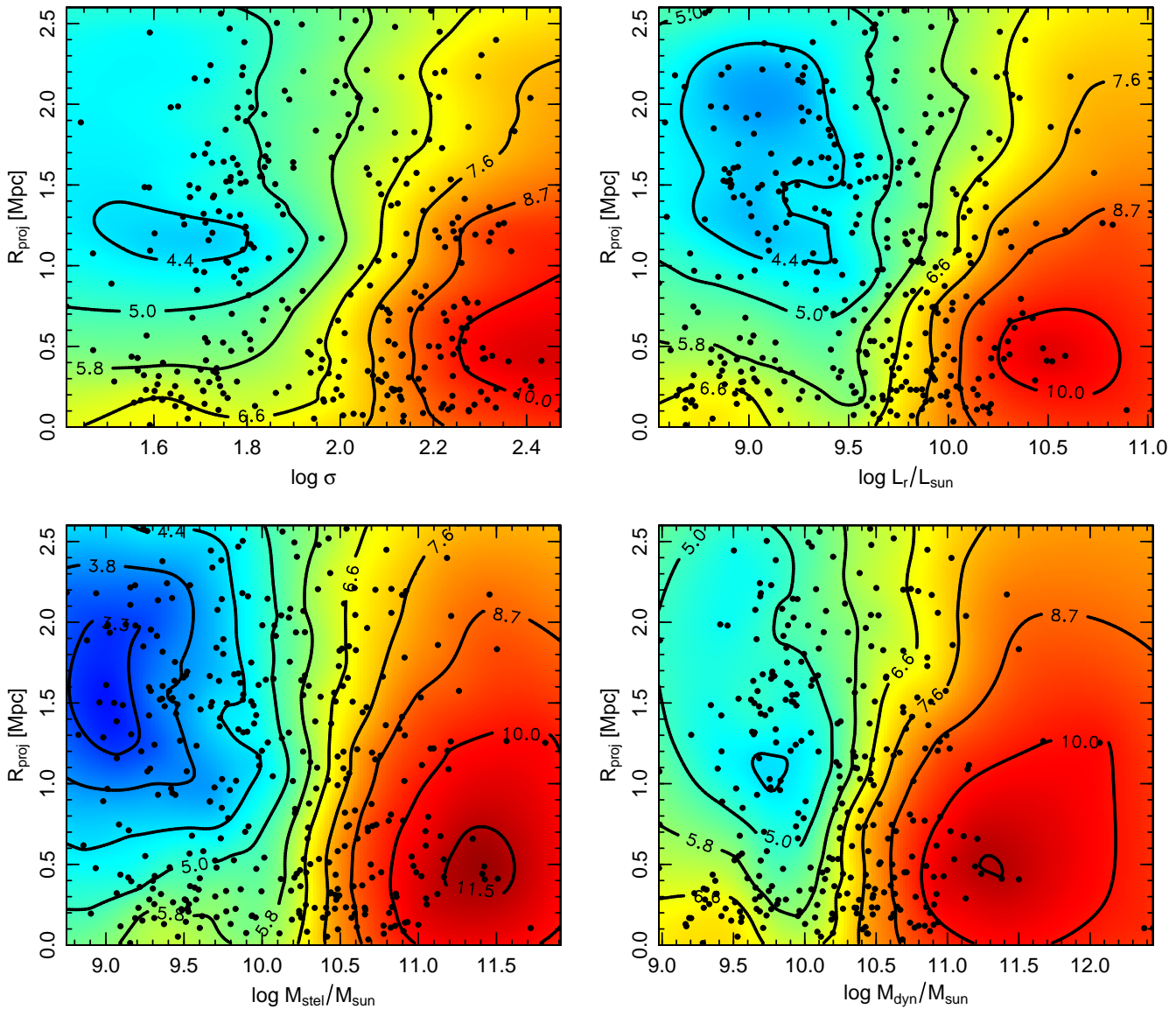}
\caption{Contour maps of average age as a function of ``mass'' and projected distance from the cluster centre, for four alternative choices of the mass-related quantity.
The maps were created using a Kriging model (e.g. Cressie 1993). 
Black points show the location of the sample galaxies within this plane.
Contours are drawn at intervals of 0.06\,dex (15 per cent) in age. The key results of Section~\ref{sec:envtrends} are evident
in all four panels: the ages of ``giants'' are related
primarily to ``mass'' (with some weak environmental modulation), while the ages of ``dwarfs'' depend mainly on their location in the cluster.
}
\label{fig:ageradsigmap}
\end{figure*}

\begin{figure*}
\includegraphics[angle=0,width=180mm]{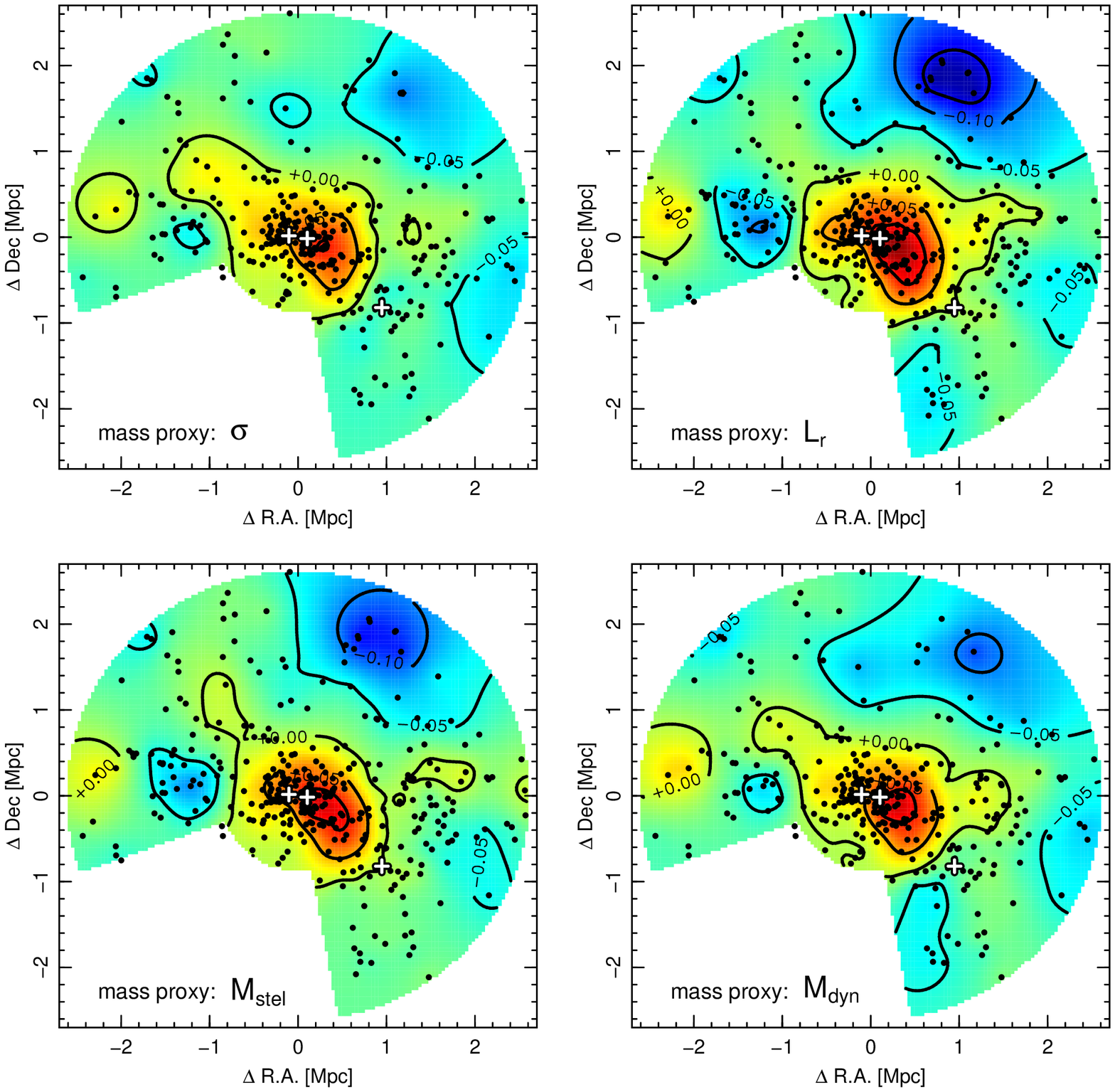}
\caption{Residuals from the age-versus-mass relations, mapped as a function of location within the cluster. Black points show the positions of the sample
galaxies, while the colour-map and contours (spaced by 0.05\,dex) give the average residual from the broken-stick age--mass relations. Each panel shows the 
residuals computed using a different mass-proxy variable as noted in the lower left. 
White crosses mark the positions of the three giant galaxies NGC 4839, NGC 4874 and NGC 4889 (from right to left). North is up and east is to the left.}
\label{fig:ageskymap}
\end{figure*}

\subsection{Observed trends of age with environment}\label{sec:envtrends}

The characteristic stellar ages of red-sequence galaxies are known to depend strongly on their ``mass'', as represented by various 
proxies such as velocity dispersion and luminosity (e.g. Caldwell, Rose \& Concannon 2003; Nelan et al. 2005; Thomas et al. 2005; 
Smith, Lucey \& Hudson 2009b), and there is a tendency for at least the most massive ellipticals to be located at the centres of clusters. 
Hence to recover secondary correlations with environment, for a sample spanning a wide mass range, it is desirable to control for the 
dependence on mass. 

In this paper, we use four different mass-related observables to remove the mass dependence and hence isolate the environmental trends: 
\begin{itemize}
\item
Central velocity dispersion, $\sigma$, is generally found to be the quantity yielding the tightest scaling relations (e.g. Smith et al. 2009b), 
but is not measured for all of the galaxies in the sample. 
\item
Luminosity, $L_r$, is measured for all galaxies, and is cleanest in terms of the sample selection criteria, but favours younger, brighter galaxies
at given stellar mass. 
\item
Estimated stellar mass, $M_{\rm stel}$, computed from $L_r$ and the spectroscopic age and metallicity, is available for all galaxies, and is in principle a more
physically meaningful quantity than luminosity.  We use the stellar mass-to-light ratio from the models of  Maraston (2005), 
assuming single-burst star-formation histories and Kroupa (2001) initial mass function. Measurement errors in age translate into $M_{\rm stel}$, hence the errors are correlated.
\item
Dynamical mass  $M_{\rm dyn}$, computed from $\sigma$ and the half-light radius $R_{\rm h}$, via $M_{\rm dyn}= 5 R_{\rm h} \sigma^2 /  G$, is an alternative physical 
indicator for galaxy mass.  This quantity does not suffer from error correlations, but is only available for those galaxies with velocity dispersion data. 
\end{itemize}

The global age--mass correlations for the full sample are shown in the upper row of Figure~\ref{fig:har_agemass}, where each panel shows results
for a different mass proxy.  As discussed in greater depth in 
S11, the sample shows an apparent break in the stellar population scaling relations, at a mass of $\sim10^{10}$\,M$_\odot$ or $\sigma\approx90$\,\kms.
The trend towards younger stellar populations in less massive galaxies tends to flatten or even reverse at low mass, as previously noted by Allanson et al. (2009). 
To capture this behaviour, we adopt a broken-stick regression model, in which a different slope is allowed on each side of a threshold mass which is itself 
a free parameter in the fit. We weight galaxies in the fit by $(\sigma_i^2+\sigma_0^2)^{-1}$ where $\sigma_i$ are the age errors on the individual data points and 
$\sigma_0$ is a constant intrinsic scatter component determined iteratively by requiring $\chi_\nu^2=1$. The errors in the mass proxies are not accounted for in the fit. 

In the subsequent rows of Figure~\ref{fig:har_agemass}, we present the age residuals from the corresponding age--mass relation in the top row, 
as a function of cluster-centric radius $R_{\rm proj}$, adopting a cluster centre mid-way between the central giant galaxies NGC 4874 and NGC 4889. 
The slopes measured using each mass proxy are summarised in Table~\ref{tab:observedslopes}.
In the second row of the figure, we show the residuals for all galaxies in the sample. A linear fit to the residuals yields a negative slope of 
0.06--0.08\,dex\,Mpc$^{-1}$. (We omit the sign in the text; all quoted gradients are negative.) 
Moreover, stronger trends are observed for the lowest-mass galaxies. 
To demonstrate this, we select and fit separately the ``dwarf'' galaxies, defined as having masses in the lowest quartile of each mass proxy,
specifically $\sigma<56\,$\kms$; L_r<1.5\times10^9\,L_\odot; M_{\rm stel}<3.4\times10^9\,M_\odot; M_{\rm dyn}<5.8\times10^9\,M_\odot$.
Note that this cut is located at a lower mass than the break in the age--mass relations.
For the dwarf subset, the recovered slopes are 0.13--0.15\,dex\,Mpc$^{-1}$, while
for ``giants'' (defined to include all galaxies above the dwarf cut, and hence still extending to fairly low masses), the gradients are significantly weaker, at 
0.04--0.07\,dex\,Mpc$^{-1}$. 
The residual trends for giants and dwarfs are shown in the third and fourth rows of  Figure~\ref{fig:har_agemass}.
These results are all robust with respect to the choice of variable used to remove the scaling with ``mass''. Indeed even fitting the gradient directly
to the measured ages {\it without} controlling for mass, we recover qualitatively similar results, albeit with increased error due to larger scatter. 

Cutting the sample further reveals some hints at the origin of the radial trend. First, limiting the fit to within $\sim$1\,Mpc steepens the environmental
dependence to 0.08--0.12\,dex\,Mpc$^{-1}$. This change can be traced to the dwarfs, which show a slope of 0.19-0.34\,dex\,Mpc$^{-1}$ inside 1\,Mpc,
though with larger error, given the restricted sample size and extent.
The slope for the giants is unchanged by restricting the radial range. 
If the south-west sector (radius 0.5--1.5$^\circ$ bounded at angles 190$^\circ$--240$^\circ$ measured east from north) covering the NGC 4839 merging group is removed, 
the recovered trends are unchanged with respect to the original result, for giants and dwarfs alike. 
Fitting instead {\it only} the core ($<1^\circ$) and the south-west sector, we recover slopes consistent with the original results for the giants and the full sample. 
For the dwarfs, however, this subset does show a slightly steeper gradient (0.15--0.20\,dex\,Mpc$^{-1}$). 
Hence while the trend is certainly not localized to the south-west, the age dependence for dwarfs might be a little steeper in the direction of the infalling group
than elsewhere.

We have seen that the relationship between age, mass and cluster-centric distance is complex, having noted at least three non-linear aspects 
to the correlations: 
(1) an apparent change in the dependence on mass, between giant and dwarf regimes; (2) steeper radial correlations in the inner part of the cluster,
and (3) stronger radial trends for dwarfs rather than giants. 
To illustrate this behaviour with minimal parametrisation, 
Figure~\ref{fig:ageradsigmap} shows smoothed maps of the average spectroscopic age in the plane defined by galaxy ``mass'' and $R_{\rm proj}$. 
The age map provides a striking visual confirmation 
of the results described above: at high mass, the near-vertical contours describe a dominant age--mass relation, modulated
by a weak radial trend, while at the low-mass end, the contours bend around to describe a relation dominated by the
cluster-centric radius dependence. As before, the key results are seen to be insensitive to which particular mass proxy is adopted. 

Finally, in Figure~\ref{fig:ageskymap}, we use the same method to map the residuals from the age--mass relationships 
as a function of position on the sky. These maps provide a way to visualise the variation of average ages with azimuthal angle, 
confirming that it is primarily the very core of the cluster, rather than the south-west region, that is ``different'' from the rest of the cluster.
In more detail, the maps suggest that the ``oldest'' region of the cluster is displaced slightly west from NGC 4874, 
and extended towards the south-west. It is tempting to speculate that this feature in the age maps might be related to the extended plume
of diffuse light in this area (Welch \& Sastry 1971; Gregg \& West 1998). For instance, perhaps this region harbours the remains of an ancient
group originally surrounding NGC 4874, which like the stellar envelope of that galaxy has been partly stripped and displaced by interactions
with the sub-cluster centred on NGC 4889.

\section{Interpretation using environmental quenching models}\label{sec:modres}

\subsection{Environmental history of cluster members in models}\label{sec:modenv}

As discussed in the Introduction, the increasing fraction of passive galaxies with increasing galaxy density or host group mass
is interpreted as evidence for environment-driven cessation or ``quenching'' of star formation. 
The results presented in Section~\ref{sec:obsres} show that for galaxies that are now on the red sequence, the time {\it since} they 
were quenched also exhibits a modulation with environment, in this case measured by cluster-centric distance. Since the events
responsible for shutting down star formation must have occurred several Gyr in the past, the {\it present} location of a galaxy may differ dramatically 
from the environment in which it was quenched. An important step in interpreting the observed trends is thus to determine how the average 
{\it environmental history} of cluster galaxies relates to their current location within the cluster.

In this section, we address this question using halo merger trees from the {\it Millennium Simulation} (Springel et al. 2005). 
We focus here explicitly on quantities related to dark-matter halo assembly (i.e. accretion into progressively larger halos, change in 
status from central to satellite within halo etc), deferring until Section~\ref{sec:quenching} any attempt to link these processes explicitly to the
 star-formation histories of galaxies. While similar investigations have been made by Berrier et al. (2009) and McGee et al. (2009) 
for the ``global'' assembly of clusters, the new element in our analysis is to address comprehensively how the environmental history is 
correlated with halo-centric radius at $z=0$.\footnote{The radial dependence of accretion time in clusters has been touched on previously by Gao et al. (2004), 
who considered only accretion of dark-matter sub-halos into the main branch of the assembling cluster, 
and by Weinmann, van den Bosch \& Pasquali (2011), who examined only the time since becoming a satellite. 
}

As analogues to the Coma cluster, we selected the five most massive $z=0$ halos from the simulation, which have total halo
masses $M_{\bf halo}=2-5\times$10$^{15}$\,M$_{\odot}$ and (one-dimensional) velocity dispersions $\sigma_v=900-1300$\,\kms. 
From these halos, we extracted the histories of all $\sim$10$^4$ model
galaxies having stellar mass $>${}$10^9$\,M$_{\odot}$ at $z=0$, as assigned by the semi-analytic model of Font et al. (2008). 
The semi-analytic machinery is only used in this step, to select galaxies of comparable final masses to those in our observed sample.
As noted by McGee et al., using stellar masses from the Bower et al. (2006) model would not yield significantly different results. The stellar mass limit
we adopt is an order of magnitude higher than the completeness limit of the simulation merger trees.

By tracing back the main progenitor branch of each galaxy, we determined the simulation time-step
 at which it first joined the ``main branch'' of the assembling cluster (i.e. the 
most massive progenitor of the cluster at that epoch). 
We also recorded the time-step at which the galaxy first became a member of a halo of mass exceeding thresholds 
 $10^{12}$\,M$_{\odot}$, $10^{13}$\,M$_{\odot}$, and $10^{14}$\,M$_{\odot}$, roughly corresponding to poor groups, rich groups, and Virgo-like clusters, respectively.
No distinction is made here between cases when a galaxy crosses these halo mass thresholds by ``joining'' a new halo, and cases in which the
host halo grows ``around'' galaxies that already belong to it.
Finally, we extracted the latest time-step at which a galaxy was a ``central'' within its own halo, and the first time-step at which it became a ``satellite'' in a separate 
halo\footnote{For some galaxies, these are not the same event. For example, a galaxy G1 that is central within a group may be accreted into a slightly 
larger group, spend some time as a satellite before merging with the central galaxy, G2, of the larger group. If the stellar mass of G1 is larger than that of 
G2, (which is quite possible given the substantial scatter in the $M_{\ \rm stel}/M_{\rm halo}$ ratio for centrals), then after the galaxies merge
G1 will become the central galaxy of the halo. At some later time, the merged halo may itself be accreted into a larger cluster, and G1 will 
again become a satellite.}.

\begin{figure*}
\includegraphics[angle=0,width=170mm,angle=0]{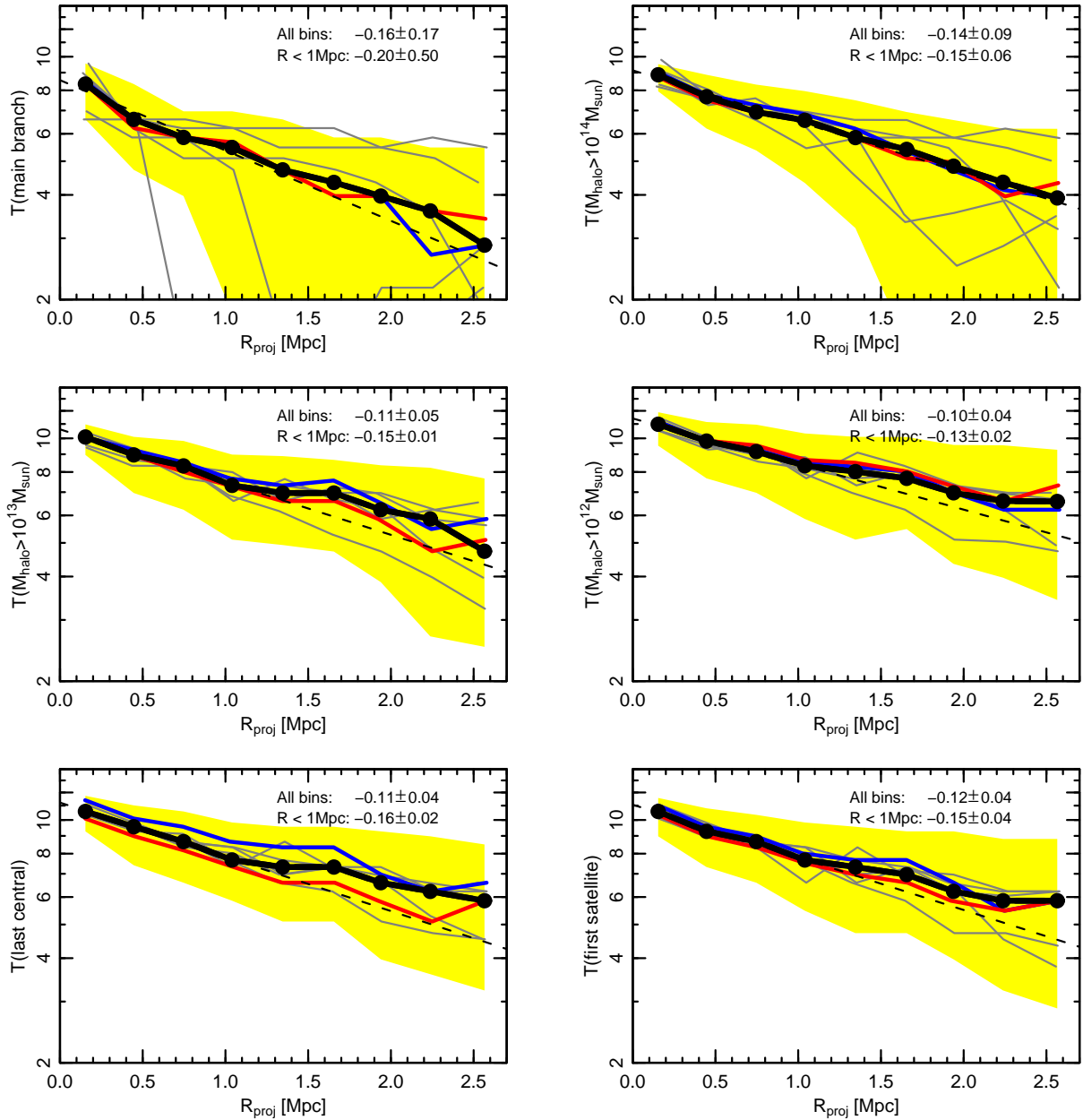}
\caption{Environmental ``ages'' of {\it Millennium Simulation} galaxies in $10^{\ 15}$\,M$_{\odot}$ halos,
as a function of projected radius from the centre of their host halo centre at $z=0$. 
Each panel shows results for a different ``event'' describing the accretion of galaxies into halos. The black points and line show the average 
time since the accretion event, in a set of radial bins, with the yellow region indicating the interquartile range computed over all galaxies in the same bins.
The thin grey lines show the results for the individual halos; the red and blue lines show the trends for galaxies in the upper and lower tercile of the 
galaxy stellar mass distribution, respectively. The legend reports the halo-centric gradient computed in two different radial ranges, 
in units of decades in age per Mpc in cluster-centric distance.  Errors here are derived from the 
range of results among individual clusters. The fit to the inner bins is also shown by a dashed line. 
}
\label{fig:envirohistory}
\end{figure*}

Having extracted these key epochs from the environmental history of model cluster members, we examine 
their correlations with projected radius from their host halo centre at $z=0$. Figure~\ref{fig:envirohistory} shows that {\it all} the events tracked 
are strongly dependent on radius, with slopes in the range 0.1--0.2\,dex\,Mpc$^{-1}$, in the sense that galaxies further out in the cluster
experienced all the events more recently than those in the core. 
The strongest radial trends are seen for the time of incorporation into the ``main branch" of the assembling halo, i.e. the final accretion event for a galaxy, and
for the time of incorporation into {\it any} halo above $10^{14}$\,M$_{\odot}$ (upper panels in Figure~\ref{fig:envirohistory}). 
(For such a high threshold halo mass, these often refer to the same event). As an example, an average galaxy observed projected near the cluster core 
first joined a Virgo-sized halo 8.9\,Gyr ago (i.e. at $z=1.3$), but at 
a projected distance 2\,Mpc, the typical galaxy encountered this cluster-like environment  only 4.8\,Gyr ago ($z=0.5$). For these events, there
is  a large scatter among the five halos studied, since they are sensitive to a small number of late mergers between high-mass halos. 

Perhaps more surprisingly, the central panels in Figure~\ref{fig:envirohistory} show that the time of accretion into {\it smaller} halos, corresponding to 
galaxy groups, is also correlated with final location in the cluster. For example,  an average galaxy observed projected near the cluster core first 
joined a $M_{\rm halo}>10^{13}$\,M$_{\odot}$ halo 10.1\,Gyr ago (i.e. at $z=1.9$), but a typical galaxy at 2\,Mpc radius encountered this group-like environment 
only 6.2\,Gyr ago ($z=0.7$).  This result can be elaborated by splitting the galaxies according to their halo mass immediately prior to accretion into the main branch. 
Some 56\,per cent of the galaxies enter the main branch without passing through any previous halo of mass greater than $10^{13}$\,M$_{\odot}$ 
(in agreement with McGee et al. 2009), so that their accretion into the main branch and first accretion into a group-scale halo refer to the same event. 
The radial trend in accretion time (into groups) for these galaxies is similar to that obtained for all objects (though by construction they are displaced to more recent accretion times). 
For the 44\,per cent of galaxies which {\it were} accreted through $M_{\rm halo}>10^{13}$\,M$_{\odot}$ halos, the accretion time (into the group) is still correlated
with location in the final cluster, although the slope of the correlation is reduced by a factor of two.
The trend of group-accretion epoch with ultimate cluster-centric radius can thus be attributed to a combination of two causes: 
(1) around half of the eventual members join the cluster without passing through a massive group stage, and 
(2) even for members which were accreted via a massive group, the time of accretion into that group was earlier among galaxies which end up in the cluster core. 
The latter occurs because galaxies in the cluster core
were formed in an over-dense region that experienced accelerated hierarchical growth, through all mass thresholds, compared to the initially more representative 
parts of the Universe which fall into the cluster at later times. 
The trend found for the lower mass threshold $10^{12}$\,M$_{\odot}$, corresponding to small groups, is probably similarly explained. 

Finally, the time at which galaxies change status, from central within their own halo to satellites within other halos, is shown by the lower panels of 
Figure~\ref{fig:envirohistory}. Note that in this case there is no threshold mass; we are simply tracking the time at which a galaxy was accreted into a halo
larger than its own. Again we find a correlation in accretion epoch with cluster-centric distance:  an average galaxy observed projected near the cluster core last ceased to be central within its own halo 10.6\,Gyr ago ($z=2.2$), but at a projected distance 2\,Mpc, the typical galaxy last ceased to be a central only 6.6\,Gyr ago ($z=0.75$). 
The slope of this correlation is comparable to that shown by Weinmann et al. (2011).

The variation in the results from cluster to cluster is substantial, particularly at large radii, 
and especially for the main-branch accretion and the $10^{14}$\,M$_{\odot}$ halo mass threshold. 
However, the overall trend of more recent ``accretion'' (however defined) at larger distance is recovered in all five clusters individually. 
The halo-centric trends are slightly stronger within the central region of the eventual cluster (within $\sim$1\,Mpc), but remain
evident at larger radius. A linear fit of $\log T$ versus $R_{\rm proj}$ provides a better description than a fit against $\log R_{\rm proj}$. 
Dividing the analysis according to galaxy stellar mass as assigned by the Font et al. model (red and blue lines in Figure~\ref{fig:envirohistory}),
we find little mass dependence in the environmental histories.
The only exception is in the time since a galaxy was last a central galaxy, which shows an offset at all radii, such that high-mass galaxies ceased to 
be centrals later than low-mass galaxies at the same projected radius. This presumably arises because galaxies continue to form stars while they 
are centrals, so that longer duration as a central galaxy leads to higher final stellar mass.

\subsection{Quenching models}\label{sec:quenching}

\begin{table*}
\caption{Observed gradients in $\Delta \log T_{\rm spec}$ (the residual from the age--mass relation) as a function of projected cluster-centric distance.
The units are decades in age per Mpc in cluster-centric distance. 
Results are given for the full sample and for various subsets,  and using four different mass proxies to define the age residual, as described in the text.
For each row, $N_{\rm \, gxy}$ gives the range in number of galaxies used in the four fits.
}\label{tab:observedslopes}
\begin{tabular}{lcccccccc}
\hline
sample & $N_{\rm gxy}$ & \ \ \ \ \ \ \  & \multicolumn{4}{c}{gradient in  $\Delta \log T_{\rm spec}$ using mass proxy} & & median gradient\\
	  &    & & $\sigma$ &  $L_r$ &  $M_{\rm stel}$ & $M_{\rm dyn}$ \\
\hline
All						& 310--362	&& $-0.06\pm0.02$  & $-0.08\pm0.02$  & $-0.07\pm0.02$ & $-0.07\pm0.02$ && $-0.07$ \\
All inside 1\,Mpc 			& 166--178	&& $-0.09\pm0.06$  & $-0.12\pm0.07$  & $-0.11\pm0.06$ & $-0.08\pm0.06$ && $-0.10$ \\
All excluding SW 			& 269--317	&& $-0.07\pm0.02$  & $-0.09\pm0.02$  & $-0.08\pm0.02$ & $-0.07\pm0.02$ && $-0.07$  \\
All in core and SW 			& 197--209	&& $-0.07\pm0.03$  & $-0.07\pm0.03$  & $-0.06\pm0.02$ & $-0.07\pm0.03$ && $-0.07$ \\
\hline
Dwarfs ($<$25\%-ile mass) 	& 78--91		&& $-0.15\pm0.03$  & $-0.13\pm0.03$  & $-0.13\pm0.04$ & $-0.13\pm0.03$ && $-0.13$ \\
Dwarfs inside 1\,Mpc 		& 40--49		&& $-0.19\pm0.13$  & $-0.20\pm0.11$  & $-0.26\pm0.14$ & $-0.34\pm0.11$ && $-0.23$ \\
Dwarfs excluding SW 		& 65--78		&& $-0.14\pm0.03$  & $-0.13\pm0.03$  & $-0.14\pm0.04$ & $-0.11\pm0.04$ && $-0.13$ \\
Dwarfs in core and SW 		& 51--57   		&& $-0.20\pm0.05$  & $-0.17\pm0.05$  & $-0.15\pm0.05$ & $-0.20\pm0.05$ && $-0.19$\\
\hline
Giants ($>$25\%-ile mass) 	& 271--232 	&& $-0.04\pm0.02$  & $-0.07\pm0.03$  & $-0.05\pm0.02$ & $-0.05\pm0.02$ && $-0.05$ \\
Giants inside 1\,Mpc 		& 118--138 	&& $-0.03\pm0.07$  & $-0.08\pm0.08$  & $-0.06\pm0.06$ & $+0.02\pm0.08$ && $-0.04$ \\
Giants excluding SW	 	& 203--243 	&& $-0.04\pm0.02$  & $-0.08\pm0.02$  & $-0.05\pm0.02$ & $-0.06\pm0.02$ && $-0.06$ \\
Giants in core and SW	 	& 140--158	&& $-0.02\pm0.03$  & $-0.04\pm0.03$  & $-0.03\pm0.03$ & $-0.01\pm0.03$ && $-0.03$ \\
\hline
\end{tabular}
\end{table*}

\begin{table*}
\caption{Age gradients predicted from quenching models applied to environmental histories of cluster galaxies in the {\it Millennium Simulation}. 
The units are decades in age per Mpc in cluster-centric distance. 
The predicted slopes are quoted for the maximally-efficient  case ($f_{\rm \ \, env}=1$) in which {\it all} galaxies are environmentally quenched. 
The quoted errors are half the difference between steepest and shallowest gradients obtained
from the five $z=0$ halos, and hence reflect expected cluster-to-cluster scatter.}\label{tab:modelslopes}
\begin{tabular}{lcccc}
\hline
quenching criterion & \multicolumn{2}{c}{--- burst or strangulation scenario ---} & \multicolumn{2}{c}{--- stripping scenario ---} \\
& $R_{\rm proj}<1$\,Mpc & $R_{\rm proj}<2.5$\,Mpc & $R_{\rm proj}<1$\,Mpc & $R_{\rm proj}<2.5$\,Mpc \\
\hline
$M_{\rm halo}$ exceeds $10^{14}$\,M$_{\odot}$ & $-0.15\pm0.06$ & $-0.14\pm0.09$ & $-0.08\pm0.03$ & $-0.06\pm0.03$ \\
$M_{\rm halo}$ exceeds $10^{13}$\,M$_{\odot}$ & $-0.15\pm0.01$ & $-0.11\pm0.05$ & $-0.08\pm0.01$ & $-0.05\pm0.02$ \\
$M_{\rm halo}$ exceeds $10^{12}$\,M$_{\odot}$ & $-0.16\pm0.02$ & $-0.11\pm0.04$ & $-0.06\pm0.01$ & $-0.04\pm0.02$ \\
\hline
Galaxy becomes a satellite for first time    	& $-0.15\pm0.04$ & $-0.12\pm0.04$ &  $-0.06\pm0.01$ & $-0.05\pm0.02$ \\
Galaxy last ceases to be a central   			& $-0.16\pm0.02$ & $-0.11\pm0.04$ &  $-0.07\pm0.01$ & $-0.05\pm0.02$ \\
\hline
\end{tabular}
\end{table*}


In Section~\ref{sec:modenv}, we examined the accretion histories of simulated galaxies and found that the timing of key accretion events remains correlated 
with projected cluster-centric distance at $z=0$. To connect this result to the trends measured in Section~\ref{sec:envtrends}, we now construct 
simple models to describe the quenching of star formation in relation to the environment, and hence predict the resulting gradients of stellar population age within clusters.

Although semi-analytic models such as Bower et al. (2006) or Font et al. (2008) provide prescriptions for the star-formation
histories of galaxies in the {\it Millennium Simulation}, they do not adequately describe all of the environmental quenching processes. 
A particular limitation, in the present context, is that even the Font et al. model,  which incorporates halo gas stripping more realistically, 
does not track the orbits of satellite galaxies within halos consistently with the dark-matter simulation. 
Instead, accreted galaxies are assigned orbital parameters randomly, from a physically-motivated parent distribution  (Section 2.2 of Font et al.). 
Since the assigned orbit dictates the stripping of halo gas, the semi-analytic star-formation histories of individual galaxies are 
not correctly related to their locations recorded from the simulation. It would therefore be misleading to compare the 
measured age gradients directly to the existing semi-analytic predictions. 

To link environmental history to predicted spectroscopic ages, we instead impose a simplistic ``quenching'' model for the star-formation history. 
The model is defined by (1) a ``quenching {\it criterion}'', i.e. the environmental conditions which 
trigger quenching and (2) a ``quenching {\it scenario}'' i.e. the form of the star-formation history, and how it relates to the time of quenching.
The quenching criterion describes the ``event'' in the environmental history of a galaxy at which the quenching is imposed. 
Following the analysis in the previous section, we consider two types of quenching criteria:
\begin{itemize}

\item {\it Halo mass threshold:}  The galaxy is quenched when it becomes part of a halo with mass above 
$10^{14}$\,M$_{\odot}$, $10^{13}$\,M$_{\odot}$, or $10^{12}$\,M$_{\odot}$ (either by accretion or halo growth). 
Hence there is an {\it absolute} halo mass scale responsible for the processes causing quenching. 

\item {\it Central-to-satellite status change:}  The galaxy is quenched when it ceases to be a central galaxy in its own halo, 
and becomes instead a satellite within another halo. In this case, quenching occurs when the galaxy falls into any larger halo, regardless
of its absolute mass. 

\end{itemize}

The above criteria account for five of the six environmental events shown in Figure~\ref{fig:envirohistory}. We do not consider the sixth case, of 
quenching only on accretion into the main branch of the assembling halo, which would imply that the ``main'' progenitor of the final cluster was in some
sense different from other halos of the same mass. This picture is clearly incorrect, for instance, in the case of late mergers between 
halos of comparable mass. 

For a chosen quenching criterion, we can extract the quenching time, $T_{\rm Q}$, for each galaxy from the simulation data. 
The quenching {\it scenario} describes the form of the star-formation history, and hence how $T_{\rm Q}$ 
maps onto the SSP-equivalent age which would be measured spectroscopically, $T_{\rm spec}$. 
We consider three types of scenario: 
\begin{itemize}

\item {\it Burst scenario:} Here we assume that quenching is caused by a star-burst which rapidly consumes all remaining cold gas in the galaxy. 
No further star formation takes place after the burst, and that the burst is sufficiently strong to dominate the ``observed'' spectrum.
Physically, this could be caused by tidal interaction with the cluster potential driving cold gas to the centre where a star-burst 
results (e.g. Byrd \& Valtonen 1990). This scenario has the advantage of simplicity, since by construction the measured SSP-equivalent age is
approximately equal to the quenching age, i.e. $T_{\rm spec} \approx T_{\rm Q}$. 

\item {\it Stripping scenario:} In this case, we assume a constant star-formation rate (SFR) beginning at very high redshift, followed by an abrupt quenching 
event at which all star formation ceases immediately. This scenario could represent a rapid ram-pressure stripping of cold gas from the disk of the infalling 
galaxy (e.g. Gunn \& Gott 1972;  Quilis, Moore \& Bower 2000). The spectroscopic (SSP-equivalent) age in this case is {\it not} equal to the quenching age, 
since all stars formed {\it before} the quenching event. However, the spectroscopic age is weighted towards the youngest 
stellar populations, both because they are more luminous, and through the non-linear age dependence of the Balmer lines (Serra \& Trager 2007). 
In S09, building on work by Allanson et al. (2009), we determined a conversion between quenching time and $T_{\rm spec}$ for the abruptly-quenched 
constant-SFR model. (A similar result was obtained by Trager, Faber \& Dressler 2008). Inverting the 
form of the relation quoted in S09 gives $\log T_{\rm spec} = \sqrt{1.087\,\log T_{\rm Q} + 0.190} - 0.07$. 

\item {\it Strangulation scenario:} The third possibility is similar to the stripping case in having a constant SFR prior to quenching, 
but now the quenching event is followed by an exponentially declining SFR, with an $e$-folding time of 1\,Gyr. Physically, this is intended to mimic 
the strangulation picture  (e.g. Balogh, Navarro \& Morris 2000) in which the hot halo gas is removed from a satellite galaxy through interaction with the 
ambient intra-halo medium, while the remaining cold gas in the galaxy disk is consumed over a longer time-scale by star formation. 
The spectroscopic age is younger than for an abrupt quenching at the same epoch, since  the post-quenching tail of
young stars compensates for the pre-quenching population. In fact, S09 showed that for  the particular case of a 1\,Gyr
decline time, the SSP-equivalent age is  approximately equal to  the quenching age, i.e. $T_{\rm spec} \approx T_{\rm Q}$. 

\end{itemize}

For the purposes of this paper, the burst and strangulation scenarios make identical predictions, since they have the same mapping from
quenching time to spectroscopic age, and we can treat them together, despite their different physical motivations. The scenarios
above are clearly simplistic, and unlikely to be consistent with other constraints on star-formation histories, but serve to bracket a wide range of
intermediate possibilities, e.g. abrupt quenching with superposed bursts. 
On physical grounds, we might expect the stripping scenario to be associated with a high halo-mass threshold criterion, since ram-pressure stripping 
of dense cold gas requires the high intra-cluster densities and velocities associated with massive clusters. 
The concentration of post-starbust spectra in the densest  parts of the Coma Supercluster, noted by Gavazzi et al. (2010) and 
Mahajan, Haines \& Raychaudhury (2011) suggests that the burst model may also be relevant in high mass halos. 
However, for completeness, we test all combinations of quenching criterion and quenching scenario in the analysis which follows. 

A final ingredient required for the models is the {\it efficiency} of environmental quenching,  $f_{\rm env}$. For maximally-efficient models, with 
 $f_{\rm env}=1$, {\it all} galaxies are quenched when they meet the chosen environmental criterion. Reducing $f_{\rm env}$ corresponds to 
assuming that some fraction of galaxies are quenched through processes which are not driven by (or even correlated with) the environmental history. 

Having chosen a quenching criterion and a quenching scenario, we can now predict the spectroscopic ages of all galaxies in the simulated clusters: 
For each galaxy we use the simulation data to identify its quenching age $T_{\rm Q}$, i.e. the time at which it met the chosen quenching threshold. Then we
convert this to a spectroscopic age using  $T_{\rm spec} = T_{\rm Q}$ (in the burst or strangulation scenario) or $\log T_{\rm spec} = \sqrt{1.087\,\log T_{\rm Q} + 0.190} - 0.07$ 
(for the stripping scenario). When $f_{\rm env}<1$, we model non-environmental quenching randomly reassigning ages for a fraction $1-f_{\rm env}$ of galaxies, drawing the new 
values from the distribution of all predicted $T_{\rm spec}$. 

As the point of contact with the observational results, we focus on the slope of the cluster-centric trends in  
spectroscopic age\footnote{The absolute age calibration of the stellar population models used to extract ages from the spectra
is uncertain at (at least) the 20\,per cent level due to errors in the stellar effective temperature scale  (Percival \& Salaris 2009). Hence comparing 
relative shifts in the ages is preferred over comparison of the ages themselves.}. The trends are fitted with respect to projected, rather than physical, 
cluster-centric distance, to reproduce the observed quantities. We fit the trends within 2.5\,Mpc and within 1.0\,Mpc, to match the 
fits made to the data in Section~\ref{sec:envtrends}.
Because the aim is to compare to observations within a single cluster, the relevant 
error on the prediction is the halo-to-halo scatter in the gradient, rather than the formal error on the fit to all halos. 
Specifically, we adopt half the range between the maximum and minimum
slope among the five halos\footnote{For a normally-distributed random variable, this quantity estimated from five samples is on average 16\,per cent 
larger than the standard deviation and is within a factor of two from the standard deviation in 90\,per cent of realisations.}.

\subsection{Results and comparison to observed trends}\label{sec:confront}

Table~\ref{tab:modelslopes} summarises the SSP-equivalent age gradients predicted by our quenching models, for the $f_{\rm env}=1$ case
(i.e. all galaxies quenched by environment-related events). 

The predicted trends within 2.5\,Mpc are in the range 0.04--0.14\,dex\,Mpc$^{-1}$, while those within the central 1\,Mpc are 0.06--0.16\,dex\,Mpc$^{-1}$.
In general, the slopes are fairly insensitive to the choice of quenching {\it criterion}, which follows from the similar slopes obtained for the 
various environmental ages in Figure~\ref{fig:envirohistory}. (Note that for the burst or strangulation scenario, the slope of the predicted ages is identical to
the slope quoted in the figure.) For the halo mass threshold quenching criteria, slightly steeper slopes are recovered for higher threshold mass. The sense of this 
effect is as expected from pre-processing: for lower threshold masses, more galaxies are quenched in groups prior to accretion into the final cluster, and
hence the slopes are diluted. The predictions depend much more sensitively on the quenching {\it scenario}, with the abrupt stripping case leading to
predicted trends that are flatter by a factor of about two. This effect is due to the different mapping from quenching time, $T_{\rm Q}$ (which is the same as in the burst
or strangulation scenarios) to SSP-equivalent age, $T_{\rm spec}$. In the centre of the halo, where most galaxies quenched early, 
this mapping has little effect, since an early short burst looks much like an SSP (e.g. $T_{\rm Q}=10$\,Gyr corresponds to $T_{\rm spec}=11.5$\,Gyr in the 
stripping scenario). However, in the outskirts, where most galaxies quenched fairly recently, the difference is much greater 
(e.g. $T_{\rm Q}=5$\,Gyr corresponds to $T_{\rm spec}=8$\,Gyr). Hence the stripping model tends to compress the range of 
SSP-equivalent age, leading to shallower predicted gradients. 

The predictions in Table~\ref{tab:modelslopes} can be directly compared to the observed gradients in Table~\ref{tab:observedslopes}.
As a first-order result, we note that the range of observed slopes 
is almost precisely spanned by the range of predictions from the quenching models. On a general level, this confirms that the observed 
slopes can be reproduced by reasonable environment-driven quenching processes. 
In greater detail, we can tentatively attempt to discriminate between the various models based on their agreement with the observed trends. 
Since the effect of reducing $f_{\rm env}$ is to dilute the predicted gradient, and tests show that the effect is roughly linear, 
i.e. setting $f_{\rm env}=0.5$ approximately halves the cluster-centric slope, a model which predicts too {\it steep} a slope in 
Table~\ref{tab:modelslopes}  can be reconciled with the data by reducing $f_{\rm env}$ accordingly. 
Only a model which predicts a slope that is too {\it shallow} compared to the data can be rejected.

For the dwarf galaxies, which show strong gradients (0.13\,dex\,Mpc$^{-1}$ within 2.5\,Mpc and 0.23\,dex\,Mpc$^{-1}$ within 1\,Mpc), the burst or strangulation
scenarios can reproduce the measured trends within the errors, so long as the efficiency is close to maximal.
By contrast the slopes predicted by the models with abrupt stripping 
(0.04--0.06\,dex\,Mpc$^{-1}$ within 2.5\,Mpc and 0.06--0.08\,dex\,Mpc$^{-1}$ within 1\,Mpc),
are shallower than observed, even for $f_{\rm env}=1$. For the $R_{\rm proj}<2.5$\,Mpc sample, the mismatch is at the $\ga$2$\sigma$ level
(for the inner sample, the observational errors are larger, and the discrepancy less significant). 
For more massive galaxies, the observed trends are much shallower (0.04--0.05\,dex\,Mpc$^{-1}$), 
and can be matched either by the stripping scenario with $f_{\rm env}>0.5$, or by the burst/strangulation scenarios with $f_{\rm env}\approx0.3-0.5$. 
Within either the burst/strangulation or the stripping scenario, the predicted gradients do not depend substantially on the environmental criterion 
adopted. Hence, using this test alone we cannot discriminate between models in which quenching is effective in group-mass halos, 
and those in which massive clusters are required.

\subsection{Caveats and limitations}\label{sec:caveats}

We highlight here a few limitations of the analysis presented above. 

\begin{itemize}

\item {\it Biases in the semi-analytic galaxy selection:} 
The models are based on semi-analytic galaxies selected at $z=0$, and hence it is implicitly assumed that 
these objects have environmental histories that are statistically similar to those of real cluster galaxies. 
From comparison to galaxy clustering statistics, there is evidence for an excess of low-mass satellite galaxies
in semi-analytic models, compared to the real universe (e.g. Li et al. 2007; Kim et al. 2009). 
If the ``missing'' satellites in real clusters have been disrupted by tidal interactions (Kim et al. 2009; Henriques \& Thomas 2010), 
then it is conceivable that the excess semi-analytic satellites are preferentially those accreted at early epochs, and having
orbits which take them close to the cluster centre. 
The expected sense of this bias would be to steepen the gradients predicted by the models, 
by preserving too many old galaxies in the cluster core. 

\item {\it Limited range of quenching criteria:}
We have not explored all possible environmental criteria for quenching. 
In particular, it is unclear that ``membership'' of a given halo, i.e. passing within its virial radius, should be sufficient to trigger quenching. 
It would be interesting to consider criteria defined by passage closer to the halo centre, where intra-cluster gas densities are much higher. 
Unfortunately, the limited number of output time-steps from the {\it Millennium Simulation} precludes an accurate 
analysis based on such an event, since the rapid passage of a galaxy through the halo core is not temporally resolved. 
It seems plausible that a quenching criterion based on a close pericentre passage would yield steeper slopes than our current models. 

\item {\it Limited range of quenching scenarios:}
Likewise, the star-formation histories assumed in our quenching scenarios are very restricted and indeed unrealistic. It would be
desirable to use pre-quenching star-formation recipes motivated by observational constraints such as the ``main sequence'' of star formation
(Noeske et al. 2007). A wider range of post-quenching behaviour should also be explored. In particular, we note that using a longer
post-quenching $e$-folding time in the strangulation models would likely result in steeper predicted gradients for a given $f_{\rm env}$.

\item {\it Inclusion of star-forming satellites:}
The observational results are limited explicitly to non-star-forming galaxies, to avoid
emission contamination of the age-sensitive Balmer lines. By contrast the simulation data implicitly include all galaxies, including those which 
may still be forming stars. As a crude test, we can remove all simulated galaxies within 1\,Gyr of their quenching time, 
at $z=0$, and re-fit the slopes. We find that the slopes computed over all bins are unchanged by excluding the youngest galaxies, while the
slopes within 1\,Mpc are flatter by $\sim$20\,per cent. Hence it seems unlikely that the inclusion of star-forming galaxies in the models produces
a serious bias. 

\end{itemize}

A full exploration of these issues is beyond the scope of this paper, but should be addressed in future work.

\section{Discussion and conclusions}\label{sec:discuss}

We have analysed the SSP-equivalent stellar ages of red-sequence galaxies in the Coma cluster, focussing on the systematic correlation of 
age with projected distance from the cluster centre. 

Combining data for dwarf and giant galaxies, from the core to the virial radius, and over a wide angular extent, has helped to clarify
many of the outstanding questions posed by S09, and previous observational work in this area. In particular it is now clear that strong radial trends in the
ages of dwarf galaxies are true, global, cluster-centric effects. The merger of the NGC 4839 group in the south-west is not the sole cause of the trends seen 
in previous work, although there is still a suggestion that the trends are somewhat stronger in this direction than elsewhere, at least for the dwarfs. 
This is qualitatively consistent with the work of Caldwell \& Rose (1997), who extended the original Caldwell et al. (1993) study to a field in the north-west 
of Coma and found a lower, but non-zero incidence of star-burst and post-star-burst galaxies as compared to the south-west. 
If the age--radius trend in Coma is not localized to a particular merging event, then we should expect to observe similar behaviour in other massive clusters. Few other systems
have been studied so intensively, in particular in the dwarf regime, where the trends are strongest. None the less, work in Abell 496 (Chilingarian et al. 2008)
and Virgo (Michielsen et al. 2008) tend to support our conclusions, albeit at lower significance due to their smaller samples and/or restricted radial coverage. 
Our work also resolves the discrepancy between the strong cluster-centric trends seen in S09 for the Coma dwarfs, and the weaker dependence found by 
Smith et al. (2006) for more massive galaxies in an ensemble of clusters. By analysing a combined sample that spans a wide range in mass within the same
cluster, we have demonstrated a transition between two regimes on the red sequence: giants for which star-formation histories are governed chiefly 
by their ``mass'' (though with some residual radius dependence) and dwarfs whose evolution depends primarily on environment. 

Many authors have recently addressed mass and environment contributions to galaxy quenching using large surveys such as SDSS, 
(e.g. Kauffmann et al. 2004; Haines et al. 2007; Peng et al. 2010), with recent advances being made through application of group/cluster catalogues to 
examine the dependence on host halo properties (e.g. Weinmann et al. 2006).
In general, these analyses employ cruder,  ``instantaneous'', measures of galaxy properties (e.g. simply whether a galaxy has been quenched or not), 
and are limited to more luminous galaxies than we probe in our study. However, such studies are able to average over large numbers of clusters
and groups over a wide mass range, and so provide a useful complement to our results from a single massive halo. 
Recently among such work, Wetzel, Tinker \& Conroy (2011) find trends of ``quenched fraction'' with mass and environment in SDSS, 
which seem to be analogous to our results. They find that a greater fraction of satellite galaxies are quenched compared to central galaxies 
in halos of all masses studied ($\sim10^{12-15}$M$_\odot$), and infer that there is no particular minimum halo mass responsible for satellite quenching. 
They find that the quenched fraction among satellites 
decreases at greater halo-centric distance in halos of all masses, but with a slope which depends on stellar mass: low-mass satellites
show a much steeper dependence than high-mass satellites.  (Note that their lowest bin is centred at  $M_{\rm stel}=8\times10^9\,$M$_\odot$, 
substantially more massive than our ``dwarfs''.) 
In a separate study using SDSS, Peng et al. (2011) highlight the apparently separable effects of ``mass quenching'' and ``environment quenching''. 
The latter, which they argue is driven solely by quenching of satellite galaxies, is apparently dependent on halo-centric radius rather than 
halo mass. 
Although these results are qualitatively similar to our conclusions in this paper, 
there is a key difference in that our stellar population ages reflect, albeit imperfectly, the time {\it since} the red satellite galaxies 
were quenched. In principle this should provide additional power to constrain the physical mechanisms responsible for driving galaxy evolution in clusters and groups. 

By directly comparing our results against predictions from simulated cluster assembly histories, we have explicitly accounted for the hierarchical growth of clusters, 
and the associated complication of pre-processing within groups prior to final infall. Our analysis shows how the typical environmental history experienced
by galaxies is correlated with their location within the cluster at $z=0$. In turn, this allowed us to test simple environmental quenching models in which 
star formation is shut down in a galaxy when it enters a halo above a given mass threshold, or becomes a satellite galaxy.

The predicted gradients are quite insensitive to what environmental ``event'' is associated with quenching the star-formation. In particular,
pre-processing in groups does not prevent age gradients from being established in the final cluster.  
In models with quenching at the group halo-mass scale, the radial dependence arises in the models partly because around half of the
galaxies fall into the cluster as central galaxies within small halos, and hence do not undergo any pre-processing. However, even among those galaxies that 
are accreted through massive groups, their history of accretion {\it into} those groups still shows some correlation with present-day location: 
location in the cluster core at $z=0$ effectively selects galaxies which passed through all stages of environmental history, including pre-processing, earlier than 
those which today reside in the cluster outskirts.

The strength of the predicted gradients does, however, depend on how the quenching event is reflected in the star-formation history. Models with very 
abrupt quenching lead to shallower age trends than those in which quenching is followed by a more gradual decline in the star-formation rate. Abrupt
stripping models are marginally ruled out for the dwarfs, but can match the shallower gradients followed by the more massive galaxies, if they operate at 
nearly maximum efficiency (i.e. if {\it all} galaxies are quenched this way, without any contribution from non-environmental causes). 
The strong trends seen for the dwarfs can be reproduced by models with a dominant burst, or an exponentially-declining star-formation rate after quenching,
though again a very high environmental quenching efficiency is required. 
These models can also match the shallower trends in the more massive galaxies, if the environmental efficiency is set lower, i.e. if many of the giants were
quenched by processes uncorrelated with halo assembly (cf. the ``mass quenching'' of Peng et al. 2010, 2011). 
This result may be related to the fact that more massive galaxies on average have a higher fraction of their mass in bulges. 
Hudson et al. (2010) showed that the colours of galaxy disks become bluer at greater cluster-centric distance, but the colours of bulges are uncorrelated
with $R_{\rm proj}$. A disk-specific quenching mechanism, as suggested by this result, would lead to lower $f_{\rm env}$ for bulge-dominated 
giants than for the on average more disky dwarfs. 

In summary, our main conclusions are:
\begin{itemize}
\item
Significant spatial trends in the SSP-equivalent ages of red-sequence galaxies are observed in the Coma cluster, such that galaxies projected
close to the cluster core are older on average than those {\it of similar mass} located further from the cluster centre. 
\item
The age trend appears to be a true global cluster-centric gradient. It is not driven by an excess of young galaxies in the south-west region, 
associated with the merging group centred on NGC 4839.
\item
The observed radial age trends are stronger for dwarfs than for more massive galaxies, and strongest in the central parts of the cluster. 
\item
Intriguingly, the oldest average ages (after controlling for mass) are concentrated $\sim$200\,kpc west-south-west of NGC 4974, in an area where significant 
diffuse stellar light has been detected. 
\item
Analysing the assembly history of massive halos in the {\it Millennium Simulation} shows that galaxies located at larger distance from the cluster centre not only 
entered the final halo later than those near the cluster core, but also passed through all key environmental thresholds at later epochs. 
\item
Simple environmental quenching models, overlaid on the accretion histories, can quantitatively 
reproduce the range of observed cluster-centric age trends. 
\item
The agreement between data and models is fairly generic and does not favour a particular halo mass threshold as responsible for stripping, nor
necessarily favour any halo mass threshold model rather than a model in which quenching follows upon simply becoming a satellite in a halo of {\it any} mass.
\item
The observed age trends for dwarfs can be reproduced by models with quenching via strangulation, but only if 
this acts with nearly maximal efficiency, i.e. almost all dwarfs were quenched by events tightly linked to their host-halo environment. 
\item
The weaker radial trends seen for more massive galaxies probably imply that 
internal processes, uncorrelated with environment, dilute the cluster-centric gradients for giants.
\end{itemize}

In this paper, we have limited our modelling analysis to interpret the cluster-centric age trends in Coma, but a more comprehensive treatment is clearly desirable. 
In addition to resolving some of the caveats noted in Section~\ref{sec:caveats}, future work should address the dependence of halo-centric age trends on
halo mass and redshift. The predictions of such models could be extended to other observables such as the passive fraction and the post-starburst galaxy fraction.
Used in combination, these constraints should provide greater power to discriminate between candidate environmental quenching schemes.

\section*{Acknowledgments}

The authors thank their colleagues in the Hectospec Coma Survey team for their contributions to assembling the dataset used here, and providing comments on 
a draft of the paper. RJS thanks Sean McGee and Vince Eke for helpful discussions about the comparison to the Millennium Simulation, and the organisers of the ESO 
workshop ``Fornax, Virgo, Coma et al.'' which stimulated publication of this work.

RJS was supported for this work by STFC Rolling Grant PP/C501568/1 ``Extragalactic Astronomy and Cosmology at Durham 2008--2013''.
JP was supported by an STFC studentship. MJH acknowledges support from an NSERC Discovery Grant.

The {\it Millennium Simulation} databases used in this paper and the web application providing online access to them were constructed as part of the activities of the German Astrophysical Virtual Observatory. 

The Kriging maps in Figures~\ref{fig:ageradsigmap} and \ref{fig:ageskymap} were created using the {\sc fields} package in the {\sc r} statistical programming
 environment ({\tt http://www.image.ucar.edu/Software/Fields}).

Funding for the SDSS and SDSS-II has been provided by the Alfred P. Sloan Foundation, the Participating Institutions, the National Science Foundation, the U.S. Department of Energy, the National Aeronautics and Space Administration, the Japanese Monbukagakusho, the Max Planck Society, and the Higher Education Funding Council for England. The SDSS Web Site is http://www.sdss.org/.
The SDSS is managed by the Astrophysical Research Consortium for the Participating Institutions. The Participating Institutions are the American Museum of Natural History, Astrophysical Institute Potsdam, University of Basel, University of Cambridge, Case Western Reserve University, University of Chicago, Drexel University, Fermilab, the Institute for Advanced Study, the Japan Participation Group, Johns Hopkins University, the Joint Institute for Nuclear Astrophysics, the Kavli Institute for Particle Astrophysics and Cosmology, the Korean Scientist Group, the Chinese Academy of Sciences (LAMOST), Los Alamos National Laboratory, the Max-Planck-Institute for Astronomy (MPIA), the Max-Planck-Institute for Astrophysics (MPA), New Mexico State University, Ohio State University, University of Pittsburgh, University of Portsmouth, Princeton University, the United States Naval Observatory, and the University of Washington.

{}

\label{lastpage}

\end{document}